\documentclass[12pt]{article}
\usepackage{graphicx}
\usepackage{color}
\usepackage{cite}
\usepackage{amsmath}

\def\css{\hbox{$c\kern0.03em(\kern-0.1em ss\kern-0.1em)$}}
\def\bss{\hbox{$b\kern0.03em(\kern-0.1em ss\kern-0.1em)$}}
\def\qss{\hbox{$q\kern0.03em(\kern-0.1em ss\kern-0.1em)$}}

\def \beq{\begin{equation}}
\def \eeq{\end{equation}}
\def\eqref#1{(\ref{#1})}
\def\bea{\begin{eqnarray}}
\def\eea{\end{eqnarray}}

\def\jpsi{J\kern-0.15em/\kern-0.15em\psi\kern0.15em}

\def\Sd{S_{\kern-0.1em ss}}
\def\SSd{{\bold S}_{\kern-0.1em ss}}
\def\Sdi{S^i_{\kern-0.1em ss}}

\def\URLtilde{\lower0.2em\hbox{$\tilde{\phantom{a}}$}}
\def\mycomm#1{\hfill\break\strut\kern-3em{\color{red}\tt ====> #1
\color{black}}\hfill\break}

\newcount\timecount
\newcount\hours \newcount\minutes  \newcount\temp \newcount\pmhours
\hours = \time
\divide\hours by 60
\temp = \hours
\multiply\temp by 60
\minutes = \time
\advance\minutes by -\temp
\def\hour{\the\hours}
\def\minute{\ifnum\minutes<10 0\the\minutes
\else\the\minutes\fi}
\def\clock{
\ifnum\hours=0 12:\minute\ AM
\else\ifnum\hours<12 \hour:\minute\ AM
\else\ifnum\hours=12 12:\minute\ PM
\else\ifnum\hours>12
\pmhours=\hours
\advance\pmhours by -12
\the\pmhours:\minute\ PM
\fi
\fi
\fi
\fi
}

\def\monthname{\relax\ifcase\month 0/\or January\or February\or
March\or April\or May\or June\or July\or August\or September\or
October\or November\or December\else\number\month/\fi}
\def\today{\monthname~\number\day, \number\year}

\def\bold#1{\boldsymbol{#1}}

\def\draft{\color{red}
$\bold{\strut\kern-3em
\hbox{\tt \Large DRAFT, NOT TO BE DISTRIBUTED:  \clock, \today.}
}$\par\noindent\color{black}}
\begin{document}
\setcounter{footnote}{1}
\vskip1.5cm
\centerline{\large \bf \boldmath EXCITED $\Omega_c$ BARYONS AS 2S STATES
\unboldmath}
\bigskip

\centerline{Marek Karliner$^a$\footnote{{\tt marek@proton.tau.ac.il}},
and Jonathan L. Rosner$^b$\footnote{{\tt rosner@hep.uchicago.edu}}}
\medskip

\centerline{$^a$ {\it School of Physics and Astronomy}}
\centerline{\it Raymond and Beverly Sackler Faculty of Exact Sciences}
\centerline{\it Tel Aviv University, Tel Aviv 69978, Israel}
\medskip

\centerline{$^b$ {\it Enrico Fermi Institute and Department of Physics}}
\centerline{\it University of Chicago, 5640 S. Ellis Avenue, Chicago, IL
60637, USA}
\bigskip
\strut

\begin{center}
ABSTRACT
\end{center}
\begin{quote}
The LHCb experiment has recently reported two excited $\Omega_c$ resonances
decaying to $\Xi_c^+ K^-$, with masses about 3185 and 3327 MeV.  We discuss
their assignment to $2S_{1/2}$ and $2S_{3/2}$ states, which can be compared
with masses based on extrapolation from the observed 1S states.  The agreement
is not perfect, but weighs against an earlier alternative assignment.
\end{quote}
\smallskip

\leftline{PACS codes: 12.39.Jh, 13.20.Jf, 13.25.Jx, 14.40.Rt}

\vfill\eject

\section{Introduction \label{sec:intro}}

The LHCb experiment has recently reported the discovery of two new
$\Omega_c^0$ resonances at $3185.1 \pm 1.7 ^{+7.4}_{-0.9} \pm 0.2$
and $3327.1 \pm 1.2 ^{+0.1}_{-1.3} \pm 0.2$ MeV \cite{LHCb:2023rtu}.
Here the errors are statistical, systematic, and based on the uncertainty
of the known $\Xi_c^+$ mass.  Five previously observed $\Omega_c^0$ states
\cite{LHCb:2017uwr,LHCb:2021ptx} were confirmed with higher statistics.  These
were interpreted as P-wave excitations of a charmed quark and an $ss$ spin-1
diquark \cite{Karliner:2017kfm}: $J^P = 1/2^-$ for $\Omega_c(3000)^0$ and
$\Omega_c(3050)^0$, $3/2^-$ for $\Omega_(3065)^0$ and $\Omega_c(3090)^0$, and
$5/2^-$ for $\Omega_c(3119)^0$, an assignment favored by lattice QCD
\cite{Padmanath:2017lng}.  A less favored picture takes the $\Omega_c(3090)^0$
and $\Omega_c(3119)^0$ as $2S_{1/2}$ and $2S_{3/2}$ \cite{Karliner:2017kfm}.

In the present paper we identify the two new resonances as $\Omega_c(3185)^0 =
2S_{1/2}$ and $\Omega_c(3327) = 2S_{3/2}$, where the subscript denotes the
total spin.  The expected 2S--1S splitting is calculated and compared with
experiment in Sec.\ II, while a similar exercise is performed for
the hyperfine splitting between the 1S and 2S states in Sec.\ III.
The choice of the favored assignment \cite{Karliner:2017kfm} whereby the five
narrow states are all taken as $1P$ is noted in Sec.\ IV, while
Sec.\ V concludes.

\section{2S--1S splitting \label{sec:1S2S}}

We are interested in the difference between 2S and 1S levels after account
has been taken of hyperfine structure.  To that end we note that in a system
of spins $s_1$ and $s_2$ and total spin $S$ the hyperfine interaction for
$s_1 = s_2 = 1/2$ is proportional to $(1/4,-3/4)$ for $s_1 = (1,0)$ while for
$s_1=1,s_2 = 1/2$ it is proportional to $(1/2,-1)$.  Thus in quarkonium ($c
\bar c, b \bar b$) systems one is interested in averages $(1/4)M(J=0)
+ (3/4)M(J=1)$ while in bound states of a spin-1/2 charmed quark and a spin-1
$\bar s \bar s$ antidiquark one is interested in averages $(1/3)M(J=1/2) +
(2/3)M(J=3/2)$.  We call these ``spin-weighted averages.''

In what follows we treat the $\Omega_c^0 = css$ states as two-body entities of
a charmed quark $c$ with mass $m_c = 1709$ MeV and a spin-1 $ss$ diquark
with $m_{ss} = 1095$ MeV \cite{Karliner:2017kfm}.  The corresponding reduced
mass, $\mu_{c,ss} = (m_c m_{ss})/(m_c + m_{ss}) = 667$ MeV,
is not far from the charmonium reduced mass $\mu_{c \bar c} = m_c/2=854.5$
MeV. With the help of the bottomonium reduced mass $\mu_{b \bar b} = m_b/2 =
2521$ MeV and a power-law extrapolation for the predicted 2S-1S difference
\beq
\Delta = \overline{2S} - \overline{1S} = E_0 \mu^p~~
\eeq
using the experimental values $\Delta_{c \bar c} = 605.3 \pm 0.3$ MeV,
$\Delta_{b \bar b} = 572.3 \pm 1.2$ MeV,
one finds $E_0 = 858.8$ MeV, $p = -0.0518$, and $\Delta_{c,ss} = 613.1$ MeV.
Here one has calculated spin-weighted averages for quarkonia with relative
weights (1/4,3/4) for $J=(1/2,3/2)$.

The observed value of $\Delta$ for the two new resonances, assuming their
assignment to $2S_{J=1/2}$ and $(2S)_{J=3/2}$ states, is based on the masses
in Table \ref{tab:omc} (1S values from Ref.\ \cite{PDG}).  
To eliminate hyperfine contributions in the $\Omega_c$ states listed in Table
\ref{tab:omc} we calculate spin-weighted averages of masses, with weight 1/3
for $J=1/2$ and 2/3 for $J=3/2$.  The observed 2S--1S
difference for the spin-weighted $\Omega_c$ states is then $(3279.8^{+2.7}
_{-1.4}-2742.3\pm1.4)$ MeV, or $(537.5^{+3.0}_{-2.0})$ MeV.  This
is to be compared with the value of 613 MeV obtained above by power-law
extrapolation from charmonium and bottomonium.

One might suspect a systematic error associated with power-law extrapolation.
Altough it is unlikely to be valid on as large a range, such an estimate
gives $\Delta_{c,ss} = 609.0$ MeV, not far from our power-law estimate.

\begin{table}
\caption{Masses of 1S and proposed 2S $\Omega_c$ resonances, in MeV
\label{tab:omc}}
\begin{center}
\begin{tabular}{c c c c} \hline \hline \\
 & $M(nS_{1/2})$ & $M(nS_{3/2})$ & $\overline{M}(nS)$ \\ \hline
1S & $2695.2\pm1.7$ & $2765.9\pm2.0$ & $2742.3\pm1.4$ \\
2S & $3185.1\pm1.7^{+7.4}_{-0.9}\pm0.2$ &
     $3327.1\pm1.2^{+0.1}_{-1.3}\pm0.2$ & 
     $3279.8^{+2.7}_{-1.4}$ \\ \hline \hline
\end{tabular}
\end{center}
\end{table}

\section{Hyperfine splitting \label{sec:HFS}}
The hyperfine splitting between the $\Omega_c^0(1S)_{1/2}$ and
$\Omega_c^0(1S)_{3/2}$, using Particle Data Group \cite{PDG} masses, is $2765.9
\pm 2.0 - 2695.2 \pm 1.7 = 70.7 \pm 2.6$ MeV.  Normally one would expect
it to be less for the 2S states (see, e.g., \cite{Quigg:1979vr}) but the
value assuming the two new states are 2S is $3327.1\pm1.2^{+0.1}_{-1.3}\pm0.2 -
[3185.1\pm1.7^{+7.4}_{-0.9}\pm0.2] = 142.0^{+2.3}_{-7.8}$ MeV.  One might
ascribe part of this difference to final-state interactions, as the two new
states have widths $50 \pm 7 ^{+10}_{-20}$ MeV ($J=1/2$ candidate) and $20
\pm 5 ^{+13}_{-1}$ MeV ($J=3/2$ candidate).  Mass shifts of the same order
as total widths can
occur.  The relative widths of the $J=1/2$ and $J=3/2$ 2S candidates
are understandable:  the $J=1/2$ state decays to $\Xi_c^+$ via an S wave, while
the $J=3/2$ state decays to $\Xi_c^+$ via a more kinematically suppressed D
wave.  If the mass shift is greater for the state with the larger total width,
it is natural to ascribe the larger-than-expected 2S hyperfine splitting mainly
to a downward shift of the $J=1/2$ state.

\section{Favored assignment of five narrow states \label{sec:allP}}
In Ref.\ \cite{Karliner:2017kfm} the favored assignment of the five narrow
$\Omega_c^0$ peaks was to the five states of a spin-1 $ss$ diquark and a
spin-1/2 charmed quark in a relative P wave.  A less likely assignment was
to take the two highest narrow peaks to be 2S, leaving two lower-mass P waves
to be found.

With higher statistics, the new LHCb data show no evidence for the lower-mass
P waves.  Furthermore, taking $\Omega_c^0(3090)$ and $\Omega_c^0(3119)$ to be
2S states would exacerbate the difference between observed and predicted
1S--2S splittings, leaving the two new states without a credible assignment.

A likely solution to both the 2S-1S splitting and the hyperfine problems is
to imagine that final-state interactions mainly lower the mass of the $J=1/2$
state, for which the final-state interactions are indeed greater, while leaving
the $J=3/2$ state mainly unshifted.  Significant deviations from naive quark
model predictions due to final-state interactions occur, for example, in the
masses of $\Lambda(1405)$ and $D_s^0(2317)$ \cite{PDG}.

\section{Conclusions \label{sec:con}}

The two new excited $\Omega_c^0$ states discovered by LHCb \cite{LHCb:2023rtu},
at 3185 and 3327 MeV, have been identified respectively as 2S$_{J=1/2}$ and
$2S_{J=3/2}$.  The 1S--2S and hyperfine splittings, though smaller and larger,
respectivly, than expected, do not deviate enough from predicted values to
jeopardize these assignments.  Confirmation of our methods may be sought in
other systems with no light quarks. The $b \bar c$ (1S,2S) system would be ideal
except only the spin-zero $B_c(1S,2S)$ masses are known, whereas only the
2S--1S mass difference is known for the $B_c^*$ spin-one states \cite{PDG,%
ATLAS:2014lga,CMS:2019uhm,LHCb:2019bem}.  A useful challenge to resolve this
question would be the detection of the soft photon in $B_c^{*+} \to B_c^+
\gamma$.

\section*{Acknowledgments}
The work of M.K. was supported in part by NSFC-ISF Grant No.\ 3423/19.


\begin{thebibliography}{99}
\def\EM{\em}

\bibitem{LHCb:2023rtu} R. Aaij {\it et al.} [LHCb Collaboration],
{\EM ``Observation of new $\Omega_c^{0}$ states decaying to the
$\Xi_c^+K^-$ final state,''} [arXiv:2302.04733 [hep-ex]].

\bibitem{LHCb:2017uwr} R.~Aaij {\it et al.} [LHCb Collaboration],
{\EM ``Observation of five new narrow $\Omega_c^0$ states decaying to
$\Xi_c^+ K^-$,"} Phys.\ Rev.\ Lett.\ \textbf{118}, 182001 (2017)
[arXiv:1703.04639 [hep-ex]].

\bibitem{LHCb:2021ptx} R.~Aaij \textit{et al.} [LHCb Collaboration],
{\EM ``Observation of excited $\Omega_c^0$ baryons in $\Omega_b^- \to \Xi_c^+
K^-\pi^-$decays,''} Phys.\ Rev.\ D \textbf{104}, L091102 (2021)
[arXiv:2107.03419 [hep-ex]].

\bibitem{Karliner:2017kfm} M.~Karliner and J.~L.~Rosner,
{\EM ``Very narrow excited $\Omega_c$ baryons,''}
Phys.\ Rev.\ D \textbf{95}, 114012 (2017) [arXiv:1703.07774 [hep-ph]].

\bibitem{Padmanath:2017lng} M.~Padmanath and N.~Mathur,
{\EM ``Quantum Numbers of Recently Discovered $\Omega^{0}_{c}$ Baryons from
Lattice QCD,''} Phys.\ Rev.\ Lett.\ \textbf{119}, 042001 (2017)
[arXiv:1704.00259 [hep-ph]].

\bibitem{PDG} 
R.~L.~Workman \textit{et al.} [Particle Data Group],
{\EM ``Review of Particle Physics,''}
PTEP \textbf{2022}, 083C01 (2022).

\bibitem{Quigg:1979vr} C.~Quigg and J.~L.~Rosner,
{\EM ``Quantum Mechanics with Applications to Quarkonium,''}
Phys.\ Rept.\ \textbf{56}, 167-235 (1979).

\bibitem{ATLAS:2014lga} G. Aad {\it et al.} [ATLAS Collaboration],
{\EM ``Observation of an Excited $B_c^+$ Meson State with the ATLAS Detector,"}
Phys.\ Rev.\ Lett.\ \textbf{113}, 212004 (2014) [arXiv:1407.1032 [hep-ex]].

\bibitem{CMS:2019uhm} A.~M.~Sirunyan \textit{et al.} [CMS Collaboration],
{\EM ``Observation of Two Excited B$^+_\mathrm{c}$ States and Measurement of
the B$^+_\mathrm{c}$(2S) Mass in pp Collisions at $\sqrt{s} =$ 13 TeV,''}
Phys.\ Rev.\ Lett.\ \textbf{122}, 132001 (2019) [[arXiv:1902.00571 [hep-ex]].

\bibitem{LHCb:2019bem} R.~Aaij \textit{et al.} [LHCb Collaboration],
{\EM ``Observation of an excited $B_c^+$ state,''}
Phys.\ Rev.\ Lett.\ \textbf{122}, 232001 (2019) [arXiv:1904.00081 [hep-ex]].

\end{thebibliography}
\end{document}